\documentclass[aps,prl,twocolumn,showpacs]{revtex4-1}
\usepackage{graphicx}
\usepackage{epsfig}
\usepackage{amsmath}
\usepackage{amssymb}
\usepackage{figsize}
\usepackage{subfigure}

\def\be{\begin{equation}}
\def\ee{\end{equation}}
\def\bea{\begin{eqnarray}}
\def\eea{\end{eqnarray}}
\def\nl{\\ \noindent}
\def\nn{\\ \nonumber}

\begin{document}

\title{The effect of amplitude fluctuations on the BKT transition}

\author{Amir Erez and Yigal Meir}
\affiliation{
 Physics Department, Ben-Gurion University of the Negev, Beer Sheva 84105, Israel}
\date{\today}

\begin{abstract}
\noindent
The Berezinskii-Kosterlitz-Thouless (BKT) transition is a generic transition describing the loss of coherence in two dimensional systems, and has been invoked, for example, to describe the superconductor-insulator transition in thin films. However, recent experiments have shown that the BKT transition, driven by phase fluctuations, is not sufficient to describe the loss of superconducting order, and amplitude fluctuations have also to be taken into account. The standard models that are extensively used to model two-dimensional superconductors are the Hubbard and XY models. Whereas the XY model allows only phase  fluctuations, the Hubbard model has an extra degree of freedom: amplitude  fluctuations. In this paper we compare two Hubbard models with the same critical temperature but with different interaction, and deduce the role of the amplitude  fluctuations in the superconducting transition. For this purpose, a novel approximation is presented and used. We derive an effective phase-only (XY) Hamiltonian, incorporating amplitude fluctuations in an explicit temperature dependence of the phase rigidity. We study the relation between amplitude fluctuations and coupling strength. Our results support existing claims about the suppression of phase rigidity due to amplitude fluctuations not present in the XY model.
\end{abstract} 


\maketitle
\noindent

\section{Introduction}
 In two dimensions, in accordance with the Mermin-Wagner \cite{Mermin1966} theorem, there could be no spontaneous symmetry breaking associated with a continuous order parameter. However, it has been shown by Berezinskii \cite{Berezinskii1972}, and by Kosterlitz and Thouless \cite{Kosterlitz1973,Kosterlitz1974}, that there could be a transition from an exponentially decaying order parameter to a power-law decaying one, as temperature is lowered through the critical BKT temperature, $T_c$.  This phase driven transition was argued \cite{Beasley1979} to be relevant to the superconductor-insulator transition in disordered thin superconducting films \cite{Goldman2010},  and even to part of the phase diagram of high $T_c$ superconductors \cite{Kivelson1995}.

A standard model to study the BKT transition in two-dimensions is the XY model, that allows only phase fluctuations. This model can be derived from the negative-$U$ Hubbard model (both models will be defined below), in the limit of large $U$, where fluctuations in the pairing amplitude can be neglected \cite{Auerbach1994}. However,  a recent experiment \cite{Yong2013} showed a measurable discrepancy between the observations and a theory that takes into account only phase fluctuations. In this case, one may expect that the full Hubbard model, that does include amplitude fluctuations, should be more appropriate to describe the physical system.  In this paper, we explore the negative-$U$ Hubbard model, and compare its predictions to that of the XY model. We concentrate on two specific values of $U$ that give rise to the same critical temperature, one where amplitude fluctuations are expected to be negligible (large $U$) and one where they have a substantial effect on the physics (small $U$). We show that one can characterize the system by an effective local Josephson coupling, and study the behavior of this effective coupling as a function of temperature for different values of $U$. This quantity, which can be probed using a local measurement, highlights the effects of amplitude fluctuations for small $U$, compared to the large-$U$ system, that is very well described by the XY model.

\section{Models and methods}
\begin{figure}[]
 \centering
  \subfigure{\includegraphics[clip,width=0.49\hsize]{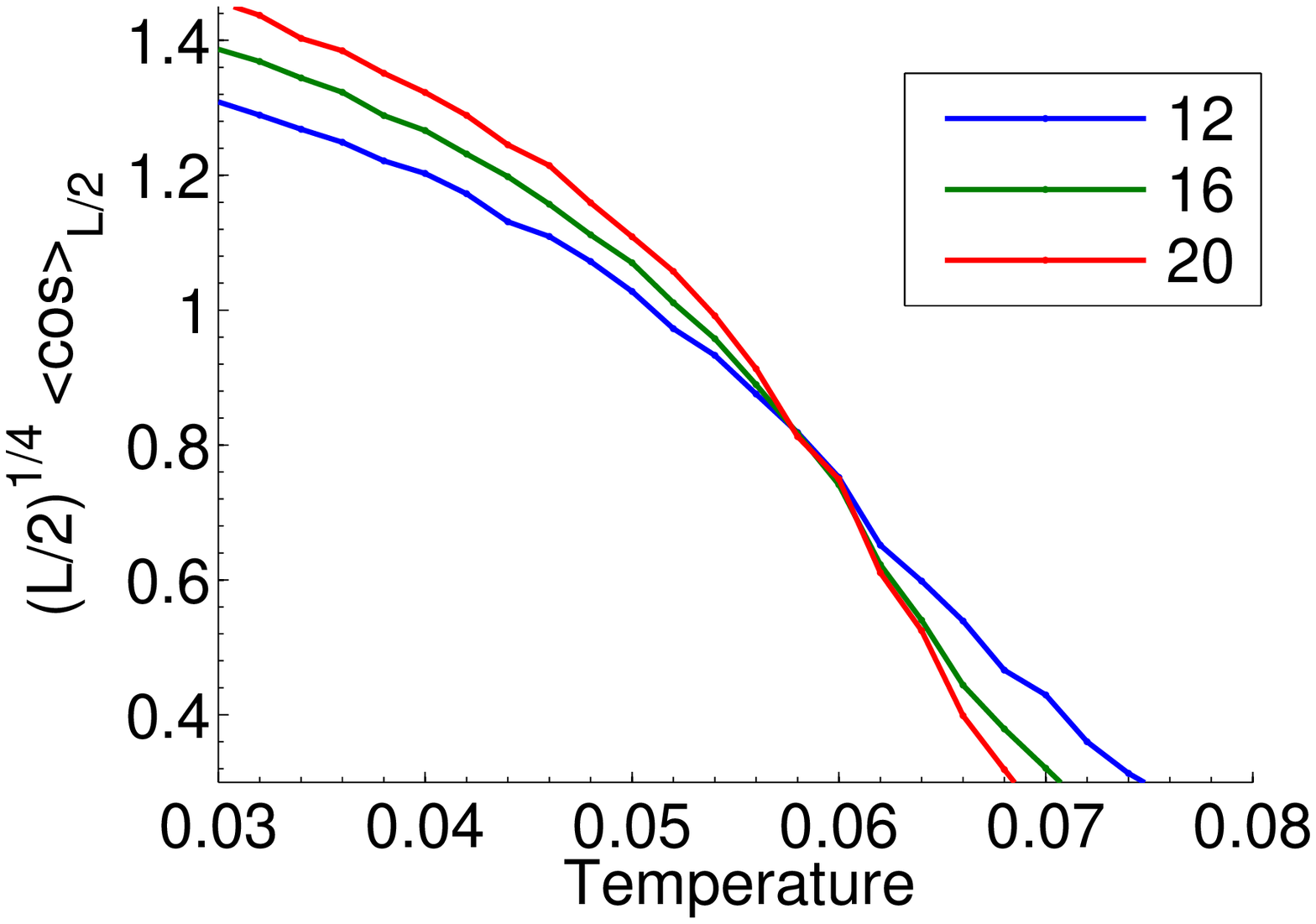}}
  \subfigure{\includegraphics[clip,width=0.49\hsize]{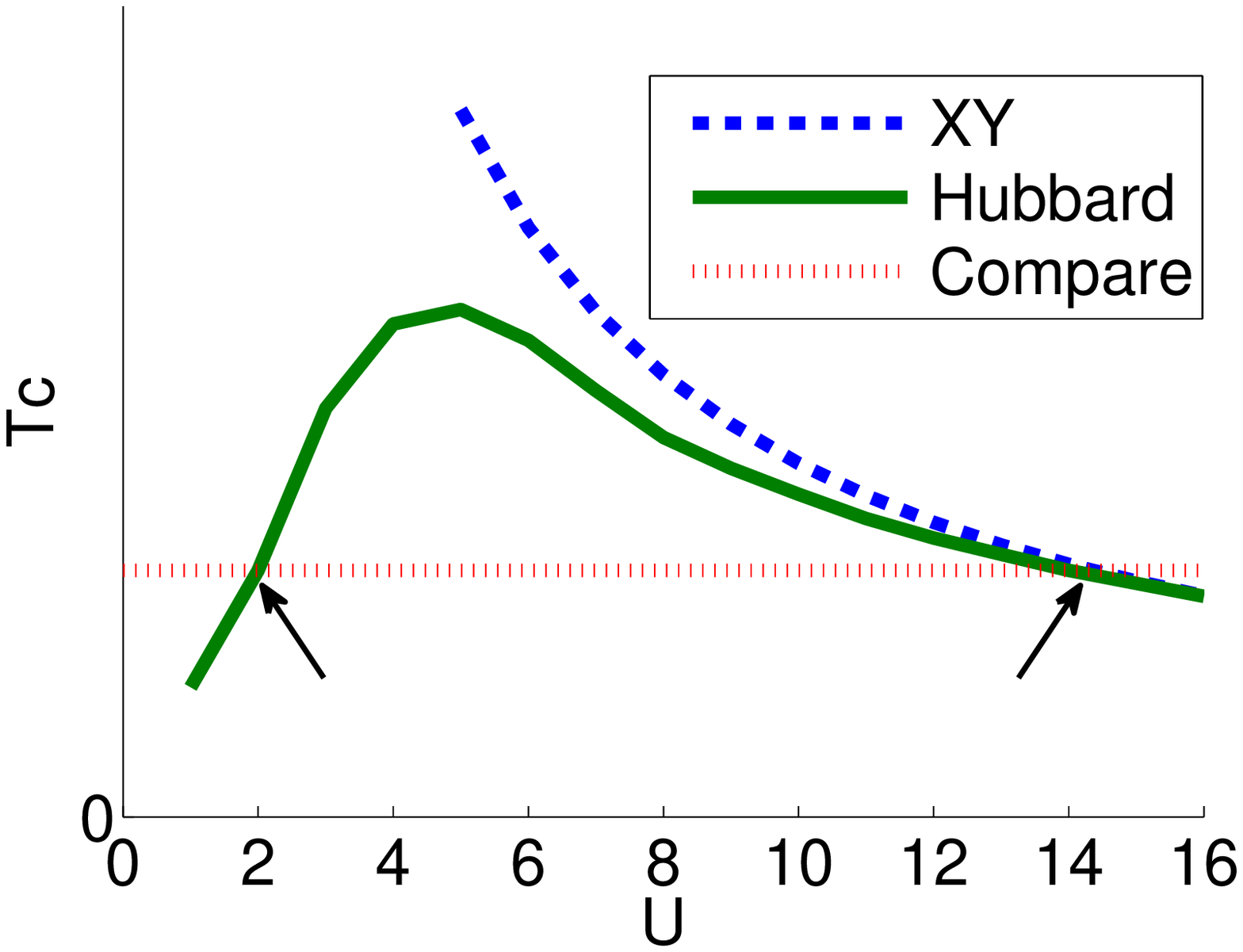}}
  \caption{\textbf{(a)} Deducing $T_c$ from the scaling relation (Eq. \ref{eq:scaling}). Scaled edge-to-edge correlation functions $L^{1/4}\langle cos(\theta_i - \theta_{i+L/2}) \rangle$ for three different size systems cross at $T_c\approx 0.058$. \textbf{(b)} Schematic drawing of $T_c(U)$ for the Hubbard model. Dashed (blue) line is the XY model $T_c \approx 0.89J$ with $J\propto \frac{t^2}{U}$. Dotted (red) line shows the two values, $U=2,15$ (black arrows) having the same $T_c$, studied in this paper.}
  \label{fig:hubbard_dome}
\end{figure}

The Hamiltonian of the classical XY-model is given by
\begin{equation}
\label{eq:H_XY}
\mathcal{H}_{\text{XY}}=-\sum_{\langle i,j \rangle} J_{ij} \cos(\theta_i - \theta_j)
\end{equation}
where $\langle i,j\rangle$ designates near-neighbors and $J_{ij}=J$ the \emph{bare} near-neighbor coupling which we take as uniform in this paper. In the context of superconductivity, the classical phases $\theta_i \in [0,2\pi)$ mimic the local phase of the superconducting order parameter. It is the thermal fluctuation of these phases which is believed to be the main mechanism for loss of conventional superconductivity in thin films and perhaps certain cuprates \cite{Kivelson1995}. 
When starting from the low temperature phase and increasing temperature towards the transition temperature $T_c$, the XY model undergoes a BKT transition at $T_c \approx 0.89J$. Usually, $T_c$ is found by use of the helicity modulus \cite{Fisher1973}, which is expected to have a universal jump (for an infinite system) at  $T_c$. However, due to the exponentially diverging correlation length at the BKT transition, a numerical 
calculation of the helicity modulus always suffers from finite size effects.  An alternative technique \cite{Paiva2004} relies on the expected scaling of the correlations 
\begin{equation}
\langle \cos(\theta_i-\theta_{i+L}) \rangle \sim \frac{1}{L^{\eta(T)}} \mathcal{F}(\frac{L}{\xi}),
\label{eq:scaling}
\end{equation}
with $\eta(T)$, the temperature dependent correlation exponent and $\xi$ the (finite-size) correlation length at temperatures below the BKT transition.
Taken with the universal BKT prediction $\eta(T_c) = \frac{1}{4}$, this provides a numerically simple way to determine $T_c$, as it only requires calculation  of the (numerically more accessible) correlation function. In Fig. \ref{fig:hubbard_dome}a we demonstrate how this scaling technique is used to determine $T_c$ for the two-dimensional Hubbard model with $U=2$. Similarly, when we employed this technique for the XY model, we found $T_c \approx 0.89J$ to good accuracy, even when using only small sized systems.\nl
Although the phase fluctuations and resulting BKT transition are well captured by the classical XY model, in real superconductors the amplitude of the superconducting order parameter can also fluctuate. Such amplitude fluctuations, physically a breaking of a cooper pair into two quasi-particles, cause suppression of $T_c$ when compared with the purely phase (XY) scenario \cite{Yong2013}. To allow such amplitude fluctuations, we use the negative-U Hubbard model,
\begin{eqnarray}
\label{eq:H_Hubbard}
\mathcal{H}_{\text{Hub}}&=&-\sum_{\langle i,j \rangle,\sigma}t \,C_{i\sigma}^\dag C_{j\sigma} - U\sum_i C_{i\uparrow}^\dag C_{i\downarrow}^\dag C_{i\downarrow}C_{i\uparrow}\nonumber\\ &-&\sum_{i,\sigma} \mu_0 C_{i\sigma}^\dag C_{i\sigma} ,
\end{eqnarray}
where $\langle i,j \rangle$ indicates a sum over nearest neighbors, $C_{i\sigma}^\dag$ creates a spin-$\sigma$ electron at site $i$; $t$ is the homogeneous hopping integral, taken to be the unit of energy, and $-U<0$ is the on-site attractive interaction. The chemical potential $\mu_0$ determines the average density $n$ and is fixed self consistently so that the density $n=0.875$ remains temperature independent.\nl
To simulate the model without resorting to quantum Monte Carlo, we use a well-established method that takes into account thermal fluctuations but ignores the quantum ones. Since we wish to simulate our systems far from a (possible) quantum phase transition, we are justified in making this approximation, thus allowing us to enjoy the relative ease of a classical simulation. Our simulation technique is explained in greater detail elsewhere \cite{Dubi2007,ErezMeir2010}. We provide here a brief description for the sake of completeness and notation.\nl
Applying a Hubbard-Stratonovic transformation to the Hubbard Hamiltonian (\ref{eq:H_Hubbard}), with a local complex Hubbard-Stratonovic field, $\Delta_i$, and ignoring the temporal dependence of these fields (quantum fluctuations), the partition function becomes:
\begin{equation}
\label{eq:partition}
Z = Tr\left[ e^{-\beta \mathcal{H}_{\text{Hub}}} \right] = \int \mathcal{D}(\{\Delta_i,\Delta_i^*\}) Tr_f [e^{-\beta \mathcal{H}_{BdG}(\{\Delta_i\})}],
\end{equation}
with the Bogoliubov-de Gennes Hamiltonian \cite{DeGennes1999} $\mathcal{H}_{BdG}(\{\Delta_i\})$ given by
\begin{eqnarray}
\label{eq:H_BdG}
\mathcal{H}_{BdG} &=& -\sum_{\langle i,j \rangle,\sigma}t\, C_{i\sigma}^\dag C_{j\sigma} -\sum_{i\sigma} \mu_0 C_{i\sigma}^\dag C_{i\sigma} \nonumber \\
 &+&  \sum_{i\sigma} \left( U_i C_{i\sigma}^\dag C_{i\sigma} + \Delta_i C_{i\uparrow}^\dag C_{i\downarrow}^\dag + \Delta_i^*C_{i\downarrow}C_{i\uparrow} \right).
\end{eqnarray}
Here $Tr_f$ traces the fermionic degrees of freedom over the single-body Hamiltonian $\mathcal{H}_{BdG}$ and can be evaluated exactly using its eigenvalues. Direct diagonalization of the Hamiltonian at each Monte-Carlo (MC) step is extremely time consuming and instead since MC updates are local, one can use a Chebyshev polynonmial expansion \cite{Weisse2006}. The integral over the fields $\{\Delta_i,\Delta_i^*\}$ can then be calculated using the (classical) Metropolis Monte-Carlo (MC) technique. One should note that unlike the BdG approximation, which amounts to a saddle-point approximation of the integral, here $\Delta_i$ are auxiliary fields. Except at zero temperature where our approach coincides with the BdG solution, the fields $\Delta_i$ are generally different from the local superconducting order parameter $<C_{i\downarrow}C_{i\uparrow}>$.\nl

Using the scaling relation (Eq. \ref{eq:scaling}), we determine $T_c$ as a function of the Hubbard $U$ for a range of values. This technique, as well as other techniques, were previously employed for finding $T_c$ of the Hubbard model \cite{Singer1998,Denteneer1993,Paiva2004,Dupuis2004}. The resulting $T_c(U)$ is schematically plotted in Fig. \ref{fig:hubbard_dome}b, and reveals a dome-like shape.  For very large values of the Hubbard coupling $U \gg t$, the kinetic term $t$ can be treated as a perturbation and a leading order expansion (away from half-filling) gives an \emph{effective XY model} \cite{Auerbach1994} with $J \propto \frac{t^2}{U}$. The proportionality factor depends on the electron density through the chemical potential $\mu_0$ and is fixed so that $T_c$ of the effective XY model matches the Hubbard one, with $J \approx \frac{t^2}{U}\,n\left(2-n\right)\,\frac{1}{1.08}$ \cite{Lindner2010}. (The last factor is due to the difference between the classical and quantum models).\nl
Note that here $J$ is temperature-independent, as is the case in the familiar XY model. The value for $T_c$ for the XY model is plotted as a dashed line which agrees with the Hubbard  critical temperature for $U>12$. Indeed, in the following section, we will explicitly show that in the large $U$ limit, amplitude fluctuations are suppressed due to the high energy cost associated with breaking a cooper pair. Thus, in the large $U$ limit, the fermionic degrees of freedom in the Hubbard model become effectively frozen and the model becomes one of interacting bosons on a lattice, ie. an XY model, with only the phases free to fluctuate.\nl


\section{Results}
Having established the shape of the Hubbard dome, we see that a given $T_c$, below the maximal possible $T_c$, corresponds to two systems with different values of Hubbard $U$. This non-trivial situation leads us to the main question we wish to address in this paper, namely, what is the difference between the low-$U$ and high-$U$ Hubbard models that share the same $T_c$ ?\nl

We chose to concentrate on two values of $U$, $U=2$ and $U=15$, both leading to $T_c \approx 0.058t$, shown in Fig. \ref{fig:hubbard_dome}b as the dotted horizontal line (for readability, let us refer to them as ’weak’ and ’strong’ coupling). The advantage of the choice of these values of $U$ is two-fold: the strong coupling side is well-approximated by the equivalent XY model; the weak coupling side still maintains a (relatively) small superconducting coherence length $\xi\approx \frac{\hbar v_F}{\Delta} \approx  5$ lattice sites. The coherence length measures the effective size of an amplitude excitation. Therefore a large coherence length necessitates simulation of larger lattice.  We find that for the parameters we chose and temperatures $T<T_c$, finite size effects are negligible. To demonstrate  this important point we have simulated two system sizes ($12 \times 12$ and $20 \times 20$) which indeed behave very similarly for $T<T_c$. In what follows, we present results for both system sizes, and compare them to the results of the XY model. Above the critical temperature $T_c$, vortex and amplitude fluctuations proliferate, necessitating much larger system sizes. Therefore, to avoid finite size effects, we limit our investigation to the $T<T_c$ temperature range.\nl

\begin{figure}[]
 \centering
  \includegraphics[clip,width=0.98\hsize]{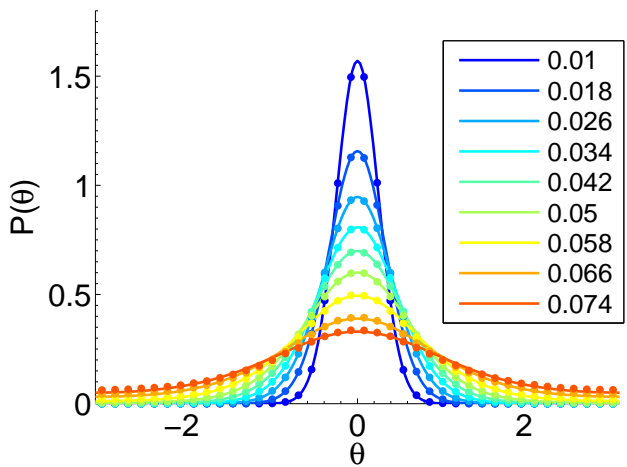}
  \caption{The distribution function of the nearest neighbor phase difference (dots) and effective Josephson fit (lines, according to Eq. \ref{eq:von_Mises}) for a range of temperatures. Results shown for a $20\times 20$ Hubbard model with $U=2$.}
  \label{fig:Hubbard_and_XY_JJ_fit}
\end{figure}

To characterize a given system, we calculate the distribution $P(\theta)$ of the phase difference, $\theta=\theta_i-\theta_j$ between two neighboring sites,  $i$ and $j$. In the absence of external magnetic field, the probability for a phase difference,  must be symmetric about $\theta=0$ with $\theta\in[-\pi,\pi)$. Since the system is invariant under translations (we use periodic boundary conditions), the distribution is identical for any nearest-neighbor pair, and we thus average it over all the pairs in the lattice. The dotted line in Fig. \ref{fig:Hubbard_and_XY_JJ_fit} shows the numerical results for $P(\theta)$ for a $U=2$ Hubbard model on a $20\times 20$ lattice for a range of temperatures. Interestingly, we find that the distribution function can be fit quite accurately by that of a single, isolated link (or Josephson junction), with an effective, temperature-dependant coupling $J_{eff}(T)$ (solid lines in Fig. \ref{fig:Hubbard_and_XY_JJ_fit}). In other words, we find
\bea 
\label{eq:von_Mises}
P(\theta) &=& \frac{1}{Z}\, e^{\beta J_{eff}(T)\cos(\theta)} \nn
Z &=& 2\pi I_0(\beta J_{eff})
\eea
with $Z$ the partition function, $\beta=\frac{1}{T}$ the inverse temperature and $I_0$ is the zeroth Bessel function.
The procedure of singling out two spins $i,j$ and integrating out all the other spins in the lattice makes the effective coupling $J_{eff}(T)$ explicitly temperature dependent. Such explicit temperature dependence resulting from partial integration has been investigated long ago \cite{Rushbrooke1940,Elcock1957} \footnote{In directional statistics, the distribution in Eq. \ref{eq:von_Mises} is known as the von-Mises distribution \cite{Mardia2009}. The von-Mises distribution is analytically convenient and is widely employed in analysing interactions in related systems (eg. liquid helium in porous media \cite{Li1990} and phase coupled neural networks \cite{Zemel1995,Cadieu2010}), making it a mathematically natural candidate as well as being physically motivated.}.\nl

 Before proceeding we wish to clarify the following:
\begin{itemize}
\item As temperature is increased, coherence between $i$ and $j$ gets decreasing contribution from other lattice sites. Indeed, at temperatures well above $T_c$ the lattice contribution to $J_{eff}$ vanishes leaving $J_{eff}$ with only the bare near-neighbor coupling. Therefore, at high temperatures in the strong coupling Hubbard model, $J_{eff} \rightarrow J \propto \frac{t^2}{U}$. Naturally, this result also applies for the XY model.
\item Amplitude fluctuations occupy a length-scale of approximately $\xi$. Therefore, to measure their effect, it is advisable to examine the coherence of near-neighbors.
\item We have made vigorous tests all showing that the above distribution also holds for spin pairs arbitrarily far apart. Moreover, the fitting function is robust for both the XY model and the Hubbard model at arbitrary $U$. 
\item The procedure we employ, integrating out part of the system, is identical to the procedure used to calculate entanglement entropy. Thus we are in the position to calculate the entanglement entropy of one pair with the rest of the lattice. The result can be analytically approximated, $S = -\int d\theta\, P(\theta) \ln P(\theta) \approx -\beta J_{eff} \frac{I_1(\beta J_{eff})}{I_0(\beta J_{eff})} + \ln 2\pi I_0(\beta J_{eff})$. We will not include further discussion of the entanglement entropy in this work. 
\end{itemize}

\begin{figure}[]
 \centering
  \includegraphics[width=0.98\hsize]{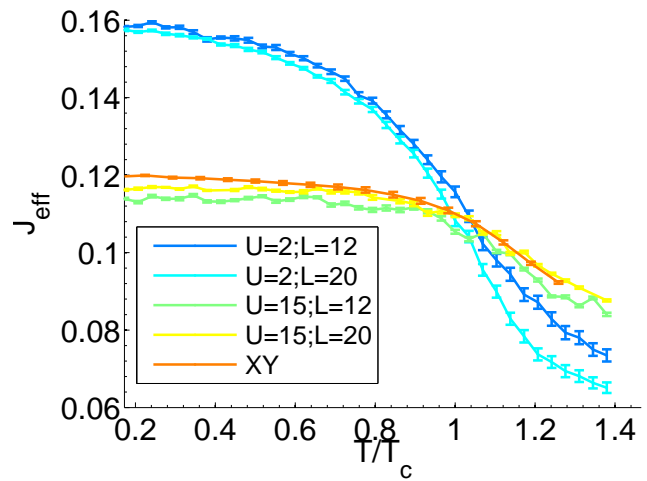}
  \caption{Effective josephson coupling $J_{eff}(T/T_c)$ extracted using the fitting procedure show in Fig. \ref{fig:Hubbard_and_XY_JJ_fit} with $T_c=0.058$. Weak ($U=2$) and strong ($U=15$) coupling Hubbard models shown (system sizes $12\times 12$ and $20\times 20$ each) as well as the $J_{eff}(T/T_c)$ for the XY model (size $20 \times 20$). The $U=15$ Hubbard behaves as an XY model, whereas for $U=2$, $J_{eff}$ is much larger at low temperatures but decreases rapidly with temperature, showing the dramatic effect of amplitude fluctuations. The two curves cross at $T_c$, consistent with the expectation that global phase coherence is lost at the same $J_{eff}$. Error bars show the fit 95\% confidence.}
  \label{fig:Hubbard_and_XY_JJ_fit2}
\end{figure}

Using this fit one can extract $J_{eff}$ which is plotted, as a function of temperature, in Fig. \ref{fig:Hubbard_and_XY_JJ_fit2}. Together, Figs. \ref{fig:Hubbard_and_XY_JJ_fit} and \ref{fig:Hubbard_and_XY_JJ_fit2} constitute the main numerical result in this paper. Fig. \ref{fig:Hubbard_and_XY_JJ_fit2} shows results for both weak and strong coupling Hubbard models in two lattice sizes, along with the XY model with the same $T_c$. Our main observation is that $J_{eff}$ for the weak coupling Hubbard is large at low temperatures but decreases rapidly with temperature, showing the dramatic suppression of near-neighbor coherence due to amplitude fluctuations. In contrast, the strong coupling Hubbard shows only a modest suppression of $J_{eff}$ for temperatures $T\le T_c$, similar to XY model. (We note that different lattice sizes produce nearly the same results demonstrating that indeed our results are insensitive to finite-size effects). These results underscore the equivalence of the strong-coupling model to the XY model, in contrast to the weak-coupling model, where amplitude fluctuations are significant.\nl
Moreover, the weak and strong coupling curves cross at $T_c$, where both systems measure the same $J_{eff}$. This agrees with an intuitive percolative description \cite{Chayes1998} of $T_c$ as the temperature which overcomes near-neighbor coupling, thus destroying the BKT order, regardless of the microscopic details of the model \cite{ErezMeir2010}.\nl

\begin{figure}
   \centering
  \subfigure[]{\includegraphics[width=0.49\hsize]{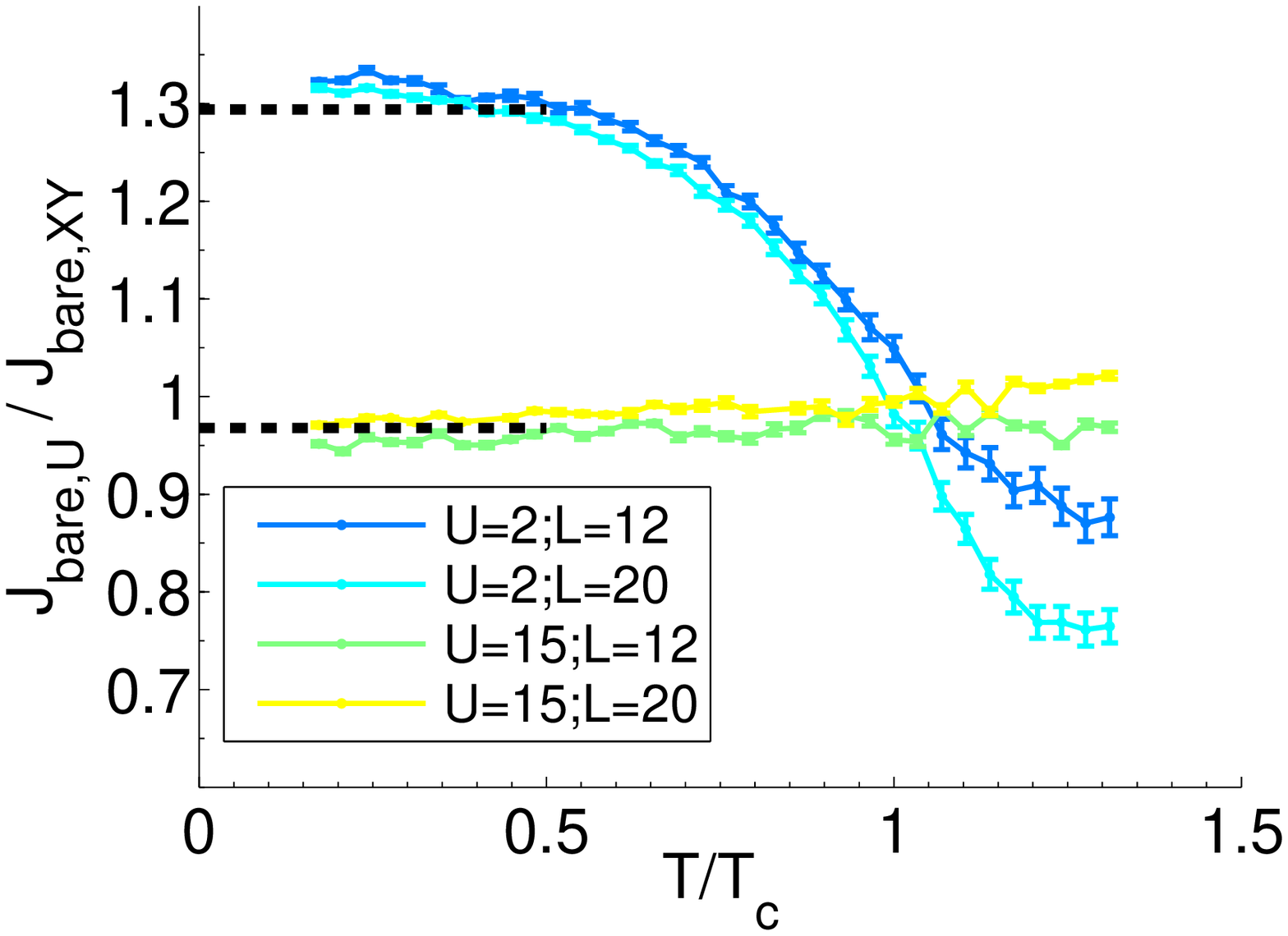}}
  \subfigure[]{\includegraphics[width=0.49\hsize]{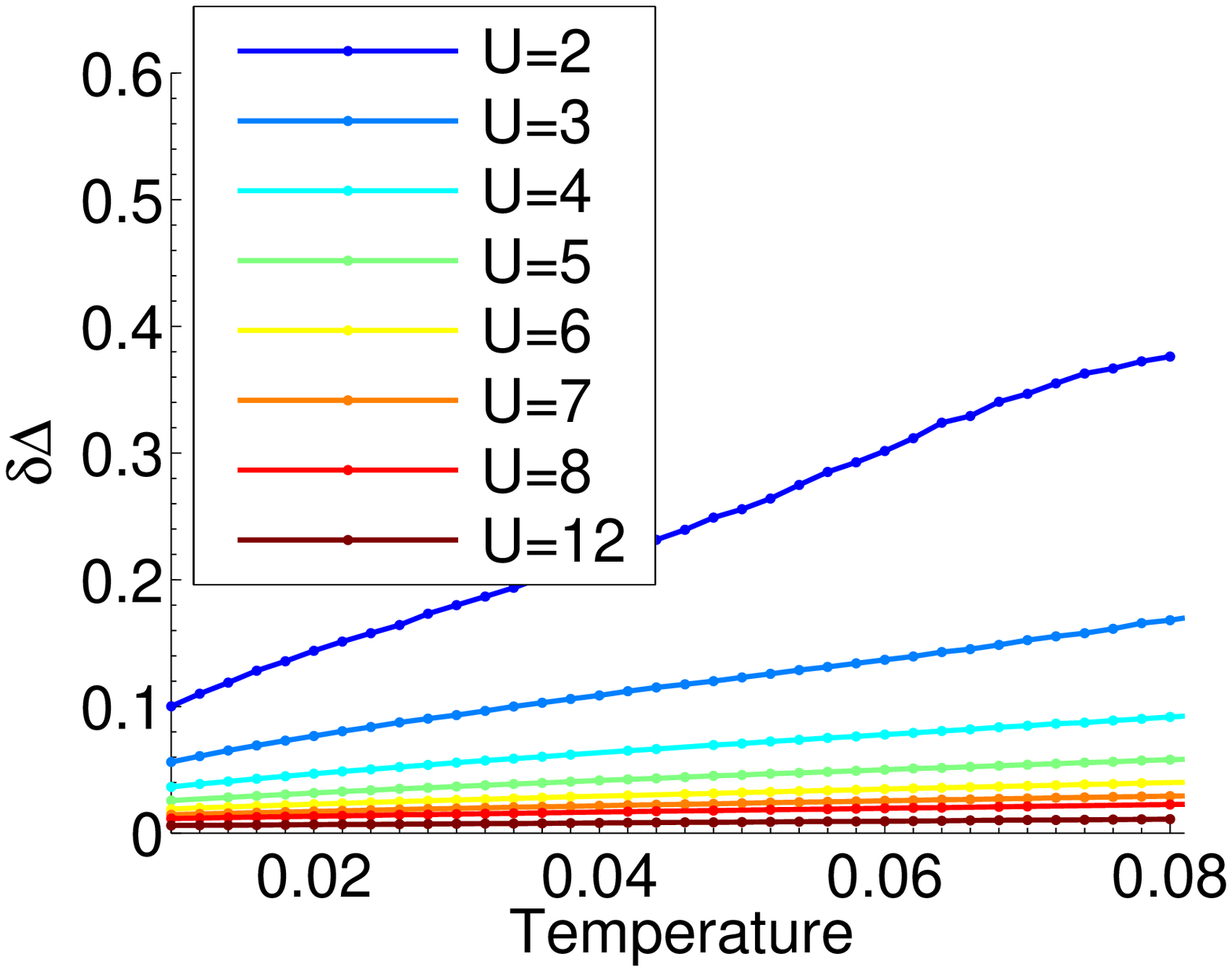}}
  \caption{\textbf{(a)} The bare coupling $J_{bare,U}$ of the XY model that gives rise to the same correlations as that of the respective Hubard model (scaled by J of the XY model with the same $T_c$, see Eq. \ref{eq:J_bare} ), as a function of T. $J_{bare,U=2}$ shows stronger temperature dependence, reflecting the effect of amplitude fluctuations. Dashed black - low temperature estimate using Eq. \ref{eq:gaussian_nn_cos}. \textbf{(b)} Relative amplitude fluctuations $\delta \Delta$ (see Eq. \ref{eq:deltaDelta}) for $U=2..12$. The weak coupling values show significant amplitude fluctuations which diminish as coupling is increased.}
  \label{fig:J_from_Jeff}
\end{figure}

In principle, since the local correlations in the negative-$U$ Hubbard model are fully captured by $J_{eff}(T)$, one can reverse-engineer a phenomenological XY model, that will lead to the same  $J_{eff}(T)$ by choosing, for each temperature, an appropriate \emph{bare} coupling $J_{bare}(T)$. Indeed, if we assume that,
\be
\frac{J_{bare,U}(T)}{J_{bare,XY}} = \frac{J_{eff,U}(T)}{J_{eff,XY}(T)}
\label{eq:J_bare}
\ee 
we can extract the \emph{bare} coupling on the LHS, using the results for the quantities on the RHS from  Fig. \ref{fig:Hubbard_and_XY_JJ_fit2}. The resulting \emph{bare} coupling ratio for both $U=2$ and $U=15$ is shown in Fig. \ref{fig:J_from_Jeff}a. For the strong-coupling case, the ratio is temperature independent; it behaves as an XY model. In contrast, the weak coupling case shows an enhanced \emph{bare} XY coupling at low temperatures which decreases significantly with temperature, to account for amplitude fluctuations \cite{Dupuis2004,Benfatto2004}.\nl

For $T \ll T_c$ it is possible to deduce $J_{bare}(T)$ by a more analytic method. Let the temperature be low enough such that for two nearest neighbors we neglect non-gaussian contributions, giving a low-energy effective Hamiltonian,
\be
\mathcal{H}_0 = \frac{J_{bare}}{2}\sum_{\langle i,j\rangle}(\theta_i - \theta_j)^2\approx \mathcal{H}_{XY}.
\ee According to the equipartition theorem, $\langle \left( \theta_i - \theta_j\right)^2\rangle_0 = \frac{T}{2J_{bare}}$.  The ensemble average $\langle ... \rangle_0$ is taken with $\mathcal{H}_0$ which means that,
\bea
\label{eq:gaussian_nn_cos}
\langle e^{i(\theta_i-\theta_j)} \rangle_0 &=& e^{-\frac{1}{2}\langle \left( \theta_i - \theta_j \right)^2\rangle_0} \nn
-\ln \langle \cos(\theta_i-\theta_j)\rangle &\approx& \frac{T}{4 J_{bare}} \quad\quad\quad \left( T \ll J_{bare} \right).
\eea
Here, $\langle \cos(\theta_i-\theta_j)\rangle$ is taken from the Monte-Carlo simulation of the full model. Thus, using the Monte-Carlo data without any additional fit, we can extract $J_{bare}(T\ll J_{bare})$ which should, naturally, agree with the $J_{bare}$ ratio we extract using Eq. \ref{eq:J_bare}.\nl
The thick dashed lines in Fig. \ref{fig:J_from_Jeff}a show the low-temperature limit for $J_{bare,U}$ derived from the procedure in Eq. \ref{eq:gaussian_nn_cos}, agreeing well with the assumption outlined in Eq. \ref{eq:J_bare}. Thus the reverse-engineering process we have suggested is consistent with its expected low-temperature limit.\nl

\begin{figure}
   \centering
  \subfigure[]{\includegraphics[width=0.49\hsize]{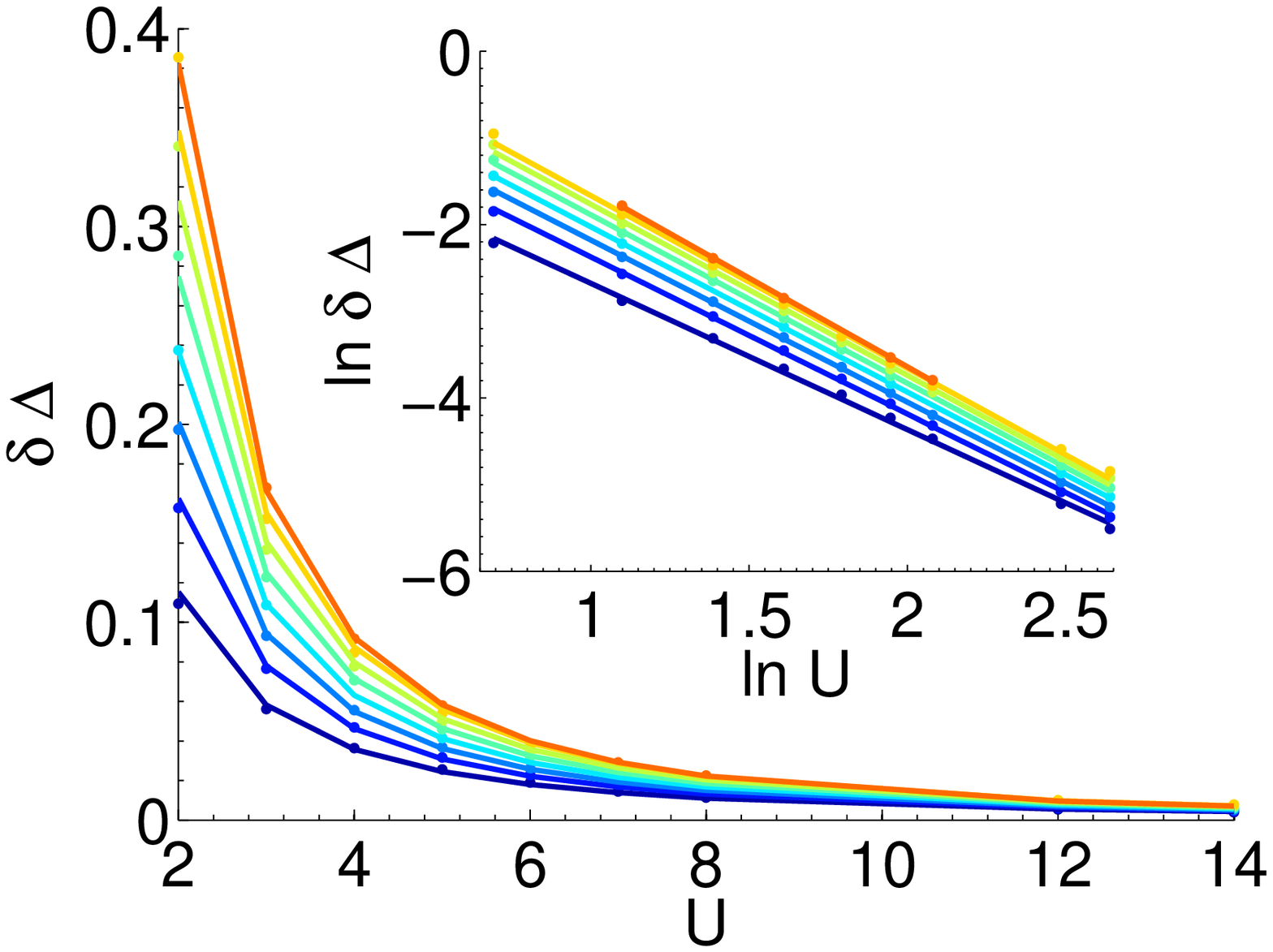}}
  \subfigure[]{\includegraphics[width=0.49\hsize]{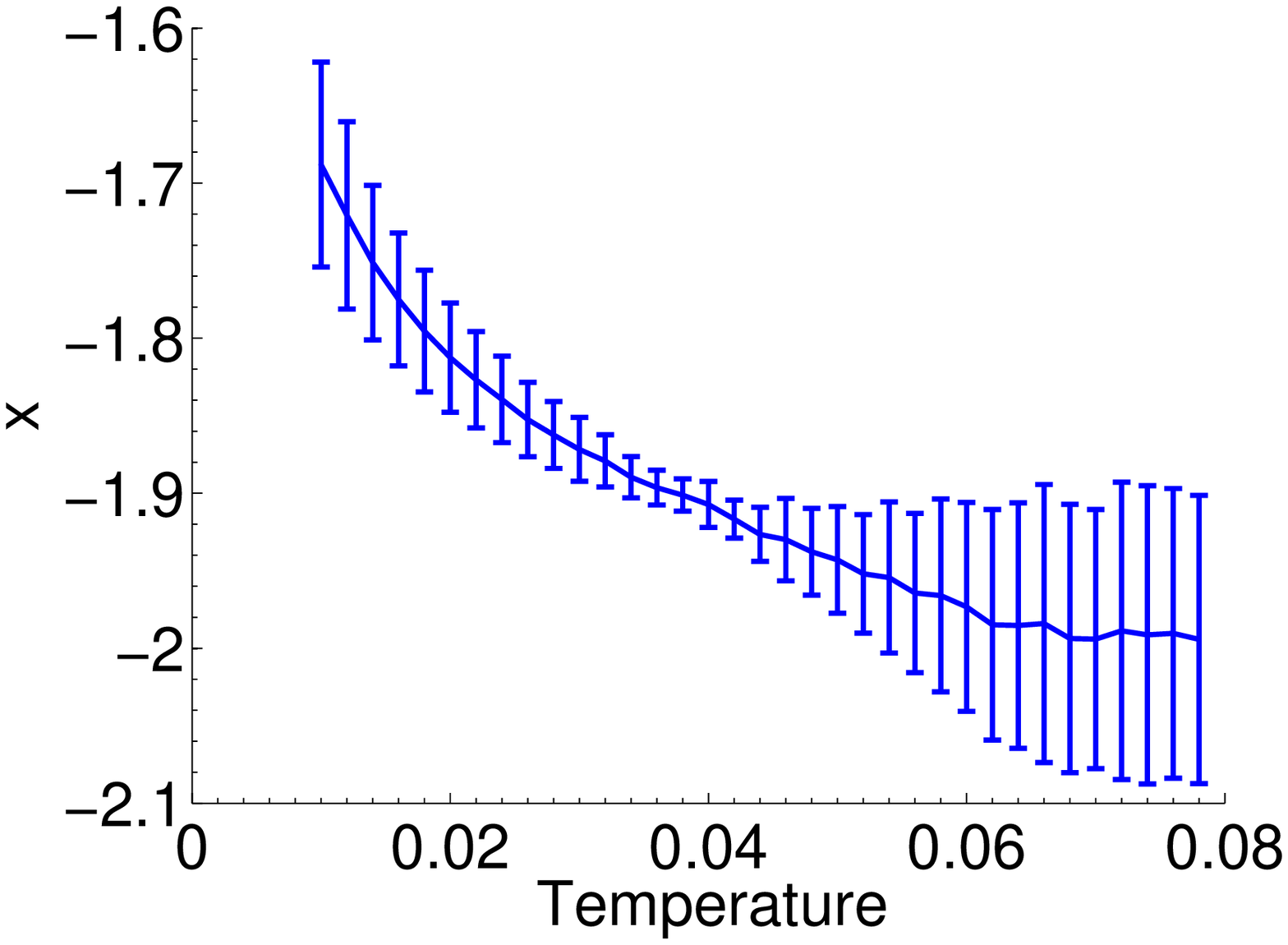}}
  \caption{\textbf{(a)} Relative amplitude fluctuations (as in Fig. \ref{fig:J_from_Jeff}b) as a function of U (dots). A power-law fit (lines) shows the relation $\delta\Delta(U,T)\sim U^{x(T)}$ at several temperatures ($T=0.01,0.02,... 0.08$ colors: blue to red). Inset - double log plot of same data (dots) and fit (lines). \textbf{(b)} $x(T)$ extracted from the power-law fit showing that $\delta\Delta$ drops faster with $U$ as temperature increases. }
  \label{fig:deltaDelta_powerlaw}
\end{figure}

So far we have discussed the effect of amplitude fluctuations without providing explicit evidence for their existence. Taking $C_{i\uparrow}$ as the (up) electron destruction operator at site $i$, we define the relative root-mean-square (RMS) amplitude fluctuations $\delta \Delta$,
\be
\label{eq:deltaDelta}
\delta \Delta(T,U) = \frac{1}{N} \sum_{i} \sqrt{\frac{\langle \vert C_{i\downarrow} C_{i\uparrow} \vert ^2 \rangle - \langle \vert C_{i\downarrow} C_{i\uparrow} \vert \rangle ^2 }{\langle \vert C_{i\downarrow} C_{i\uparrow} \vert \rangle ^2}}
\ee
With $\langle ... \rangle$ the Monte Carlo (ensemble) average, $N$ the total number of lattice sites and $\sum_i$ a sum over all lattice sites.\nl
We plot $\delta\Delta$ in Fig. \ref{fig:J_from_Jeff}b for values of $U\in[2,12]$. Clearly, significant RMS amplitude fluctuations exist only in the weak coupling case, and are, of course, stronger at higher temperatures. Fig. \ref{fig:deltaDelta_powerlaw}a shows that these curves, for the limited range of $U$ that we have calculated, satisfy a power-law relation $\delta \Delta(T,U) \sim U^{x(T)}$. The exponent $x(T)$ and its 95\% fit confidence values is plotted in Fig. \ref{fig:deltaDelta_powerlaw}b. Interestingly, $\delta\Delta(T,U)$ drops faster with $U$ for higher temperatures.

\section{Summary and Conclusions}
Realistic superconducting thin films support both phase and amplitude fluctuations of the order parameter. The simplest model that allows these two degrees of freedom is the the attractive two-dimensional Hubbard model. It is known that the critical temperature $T_c(U)$ of the Hubbard model has a dome-like dependence on the coupling $U$. Therefore, there exist two values of $U$, which we call weak ($\vert U \vert < 4$) and strong ($\vert U \vert \gg 4$), that satisfy the same $T_c$. In this paper, we have suggested a Josephson approximation that probes the phase coherence of near neighbors. The approximation is robust across a wide range of temperatures and is model-independent. Using this approximation we were able to extract an effective coupling $J_{eff}(T)$ which we use as a probe to compare the two sides of the Hubbard dome. We have found a qualitative difference between the two sides of the dome, in the temperature dependence of $J_{eff}(T)$. Specifically, in contrast to the strong coupling side, which agrees quantitatively with XY model, the weak coupling model shows a steep decrease of $J_{eff}(T)$ as temperature is increased towards $T_c$. We explain this behavior by the effect of thermal fluctuations on the superconducting amplitude and show evidence to support our explanation. 
Thus, both processes contribute to the quantitative way in which  superconductivity is lost in weak coupling films, in contrast with the simpler (phase only) mechanism in the strong coupling regime. In the regime where  amplitude fluctuations are important, our approach can be used to generate an effective phase-only action with a temperature dependent $J(T)$ as shown in Fig \ref{fig:J_from_Jeff}a. The resulting $J(T)$ can be compared against analytical attempts that integrate out the amplitude fluctuations. It can also be used when analyzing hetero-layered systems (eg. \cite{Yuli2008,Wachtel2012}) where different layers can have different coupling and therefore different amplitude fluctuations effects. Specifically, it will be interesting to use our technique to analyze bi-layered systems where both layers have the same $T_c$ but one layer is weakly coupled whereas the other strongly coupled.  
Experimentally, $J_{eff}$ could be measured by a two-tip STM experiment, perhaps by coupling the two tips to a SQUID or other sensitive device. Single-tip STM experiments have already been successfully used to probe inhomogeneous thin films \cite{Sacepe2008, Sacepe2011}.\nl

One advantage of our approach is that it readily generalizes to disordered systems. In fact, our approach is likely to apply to a wide range of systems like cuprates, heavy fermion and organic superconductors, magnetic systems as well as any coherent system where there exist both phase and non-phase degrees of freedom.\nl

\textbf{Acknowledgment:} We thank Assa Auerbach for his valuable input. This work is supported by the ISF. Amir Erez is supported by the Adams Fellowship Program of the Israel Academy of Sciences and Humanities. Most calculations were done on the BGU HPC cluster.


\bibliography{XY_vs_Hubbard}



\end{document}